# Hybridization of Epsilon-Near-Zero Modes via Resonant Tunneling in Layered Metal/Insulator Double Nanocavities

*Vincenzo Caligiuri\*[†], Milan Palei[†§], Giulia Biffi[†§] and Roman Krahne\*[†]*

[†]Istituto Italiano di Tecnologia, Via Morego 30, 16163 Genova, Italy

[§]Dipartimento di Chimica e Chimica Industriale, Università degli Studi di Genova, Via Dodecaneso, 31, 16146 Genova, Italy

KEYWORDS Epsilon-near-zero, Metal-Insulator-Metal, Resonant Tunneling, Two-Level Systems, Metamaterials, double-quantum-well

*CORRESPONDING AUTHOR roman.krahne@iit.it; vincenzo.caligiuri@iit.it

ABSTRACT

The coupling between multiple nanocavities in close vicinity leads to hybridization of their modes. Stacked Metal/Insulator/Metal (MIM) nanocavities constitute a highly versatile and very interesting model system to study and engineer such mode coupling, since they can be realized by lithography-free fabrication methods with fine control on the optical and geometrical parameters. The resonant modes of such MIM cavities are epsilon-near-zero (ENZ) resonances, which are appealing for non-linear photophysics and a variety of applications. Here we study the hybridization of ENZ resonances in MIMIM nanocavities, obtaining a very large mode splitting reaching 0.477 eV, Q factors of the order of 40 in the visible spectral range, and fine control on the resonance wavelength and mode linewidth by tuning the thickness of the dielectric and metallic layers. A semi-classical approach that analyses the MIMIM structure as a double quantum well system allows to derive the exact analytical dispersion relation of the ENZ resonances, achieving perfect agreement with numerical simulations and experiments. Interestingly, the asymmetry of the mode splitting in a symmetric MIMIM cavity is not reflected in the classical model of coupled oscillators, which can be directly related to quantum mechanical tunneling for the coupling of the two cavities. Interpreting the cavity resonances as resonant tunneling modes elucidates that can be excited without momentum matching techniques. The broad tunability of high-quality ENZ resonances together with their strong coupling efficiency makes such MIMIM cavities an ideal platform for exploring light-matter interaction, for example, by integration of quantum emitters in the dielectric layers.

INTRODUCTION

The design of photonic nanocavities with tunable resonance wavelengths is of fundamental interest for many purposes in optics. Among the important aspects for such nanocavities are their ease of integration with other photonic elements,[1] the quality factor[2,3] and tunability range of the resonance,[3,4] and the cavity losses.[5–7] Plasmonic nanocavities have the advantage of strong subwavelength light confinement and resonance tunability.[8] However, high quality factors are difficult to achieve,[9] intrinsic losses caused by the metals set a limit for light amplification, and possibly the integration with other systems is highly demanding due to elaborate three dimensional shapes of the plasmonic resonators. In this respect, nanocavities with epsilon-near-zero (ENZ) resonances are a promising alternative, since in this case the imaginary part of the effective dielectric permittivity at their resonances is small,[5] and ENZ nanocavities with possibly multiple tunable resonances with high quality factor could constitute a promising alternative as photonic resonators. Towards such tunable cavities, the coupling of resonators provides a viable approach that enables to tailor the optical response, since coupled resonators manifest a splitting of their resonant modes.[10–13]

Planar photonic nanocavities made of metal/insulator/metal (MIM) layers constitute a versatile platform for engineering nanocavities with strong light confinement in a broad frequency range[1,6,22–26,14–21], and the possibility of fabricating vertically stacked systems allows for the design of coupled resonators.[8] Furthermore, their ease in fabrication, together with the broad variety of employable materials, makes MIM cavities an ideal system for the exploration of non-linear optical properties.[11,27,28] The effective dielectric permittivity of MIM nanoresonators can be designed to manifest very high positive or negative values (eventually showing a pole),[27,29,30] or to completely vanish, manifesting ENZ behavior if losses are sufficiently small.[31–34] Such ENZ resonances have been referred to as "Ferrell-Berreman" modes, because of the similarity to the zero crossing of the dielectric permittivity in plasmon-polaritonic and phonon-polaritonic thin films.[5,35–38] However, for few-layer metal/insulator stacks the origin and nature of the ENZ resonances is still under investigation.[5,7,39] For example, the effective medium theory fails to predict their effective permittivity,[5,40,41] and it can be debated if the ENZ resonances are photonic or plasmonic, or if classical or quantum-mechanical models should be applied.[7] In a recent work, we demonstrated that MIM cavities can be seen as a double barrier quantum well, and that the ENZ cavity modes can be treated as the resonant tunneling of photons.[7] Furthermore, the occurrence of an ENZ frequency in the permittivity is usually accompanied by an increase of the local density of photonic states,[32,33] which can be exploited for the enhancement of the radiative rate of, for example, a perovskite nanocrystal film deposited on the surface of a MIMIM structure. [5]

In this paper, we study the mode splitting of the resonances in layered MIMIM double nanocavities. We demonstrate the ENZ nature of the resonant modes by spectroscopic ellipsometry, and map the anticrossing of the modes by detuning the resonance frequency of one cavity through variation of the thickness of one dielectric layer. We show that the coupling strength that governs the mode splitting can be finely tuned via the thickness of the central metal layer, and we demonstrate that the linewidth of the resonances can be optimized by acting on the thickness of the outer metal films. To gain deeper insight into the optical properties of such MIMIM cavities, we present a semi-classical analysis of their resonant modes that demonstrates that the MIMIM cavity can be viewed as a double quantum well. The symmetric and antisymmetric resonances occurring at lower and higher energy, respectively, represent the typical two-level system as it is successfully described in quantum mechanics, with the ammonia molecule as a prominent example. In particular, the MIMIM system manifests an asymmetric mode splitting, which is characteristic for quantum mechanical coupling that is non-linear, in contrast to the classical system where oscillators are linearly coupled.[10]  Other advantages of the double quantum well model for the MIMIM structure lie in the analytical solutions for the resonance frequencies, and the fact that it directly elucidates why the cavity resonances can be excited by illumination from the top, through the metal layers, without the need of a momentum matching technique. We note that the system that we address in this work is different from the one discussed by H. Alaeian and Dionne in ref.[8], since there plasmonic modes in MIMIM structures with much thinner dielectric layers were investigated that do not correspond to ENZ resonances.

RESULTS AND DISCUSSION

The structure of an experimentally fabricated MIMIM double cavity consisting of silver (Ag) as metal and $Al_2O_3$ as insulator (dielectric) is sketched in Figure 1a. Here the dielectric cavity layers have nearly similar thickness with 100 and 115 nm, and the three metal layers of Ag were chosen with a thickness of 20 nm. Such a double cavity manifests two distinct peaks in absorbance and transmittance, and two corresponding minima in reflectance that occur at 470 nm and 610 nm, and which can be associated to the high and low energy cavity modes (Figure 1b). The real ($\varepsilon'$) and imaginary ($\varepsilon$) part of the ellipsometrically measured dielectric permittivity is depicted Figure 1c and 1d, respectively. The curve of $\varepsilon'$ crosses the zero at four wavelengths, two of which ($\omega_{HE\text{-}ENZ}$ and $\omega_{LE\text{-}ENZ}$) are characterized by a small imaginary part, which renders them high-quality ENZ modes. Such ENZ resonances correspond to Ferrell-Berreman modes that occur naturally in thin Ag films (at 327 nm), and which can be designed within certain spectral bands in layered metamaterials.[5,38] Interestingly, the wavelengths of these ENZ modes coincide with the cavity features in the optical spectra in Fig. 1b, as highlighted by the vertical dashed lines, which indicates that the cavity modes of the MIMIM correspond to ENZ resonances. For an analytical modeling of the effective dielectric permittivity of the MIMIM structure, we extended the harmonic oscillator model that we developed for a single MIM system in ref [5]. Then the double cavity system can be described by adding two damped oscillator terms, one for each MIM, to the Drude expression of the dielectric permittivity of Ag:

$$\varepsilon_{eff,MIM(Ag)} = \varepsilon_\infty - \frac{\omega_p^2}{(\omega^2 + i\gamma\omega)} - \frac{\alpha_1 \omega_{MIM}^2}{\left(\omega^2 - \omega_{0,MIM1}^2 + i\gamma_{MIM1}\omega\right)} - \frac{\alpha_2 \omega_{MIM}^2}{\left(\omega^2 - \omega_{0,MIM2}^2 + i\gamma_{MIM2}\omega\right)};$$ (1)

Here $\gamma_{Ag} = 0.021\text{eV}$ and $\omega_p = 9.1$ eV are the Drude parameters of Ag. $\varepsilon_\infty$ is taken as fitting parameter and we obtain $\varepsilon_\infty = 6.8$, which is slightly larger than the value of Ag due to the residual polarizability of the system. The parameters $\omega_{0,MIM1} = 2.452\text{eV}, \omega_{0,MIM2} = 2.7382$ eV, $\gamma_{MIM1} = 0.075\text{eV}$ and $\gamma_{MIM2} = 0.07\text{eV}$, are obtained from the experimental ellipsometry spectra. Moreover, it is convenient to fix the parameter $\omega_{MIM}^2$ (at 3.53 eV) and to express the difference between the two oscillators in the numerator by a coefficient $\alpha_i$ (where $i$ is the $i^{th}$ resonance). In this case, $\alpha_1 = 0.35$ and $\alpha_2 = 0.3$. The fitting with this approach is depicted by the open circles in Figure 1c,d , and shows very good agreement with the experimental data. We note that eq. 1 allows to describe the MIMIM system as one homogenized layer with an effective dielectric permittivity.

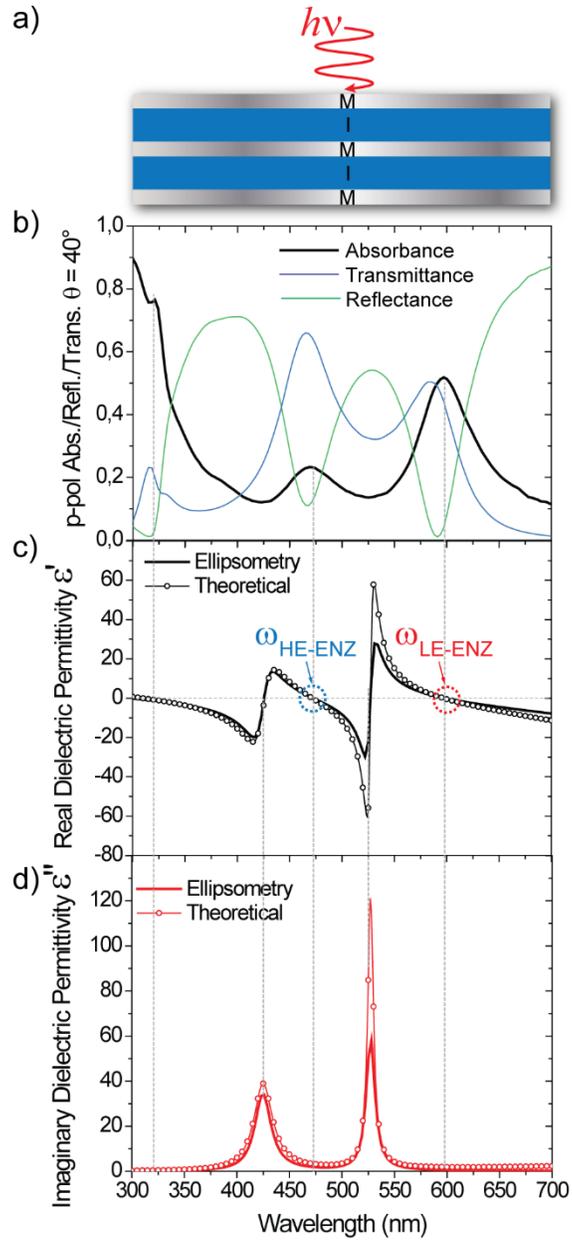

**Figure 1.** (a) Architecture of the MIMIM structure with illumination from the top. (b) Ellipsometrically measured p-polarized transmittance (blue), reflectance (green) and absorbance (black), detected at $\theta=40°$, showing two absorbance maxima at the two low-loss ENZ wavelengths. (c,d) Theoretically modeled (empty circles) and ellipsometrically measured (solid line) real (c) and imaginary (d) effective dielectric permittivity of a MIMIM cavity. Low-loss ENZ wavelengths are highlighted with a blue (high-energy) and red (low-energy) dashed circle.

Modeling of the norm of the electric field with finite element methods (COMSOL) at the resonance wavelengths shows that the cavity modes are strongly confined in the dielectric layers, as demonstrated in Figure S1 in the SI.

The MIMIM double cavity can be seen as two MIM cavities stacked on top of each other that are connected by the central metal layer. This configuration resembles two coupled oscillators, that lead to a mode splitting when the individual resonance frequencies are similar or only slightly detuned. In the case of MIM and MIMIM cavities the resonance frequencies are mainly determined by the thickness of the dielectric layers, and therefore we can control the detuning in MIMIM cavities by varying the thickness of one dielectric layer, while keeping the other one fixed.

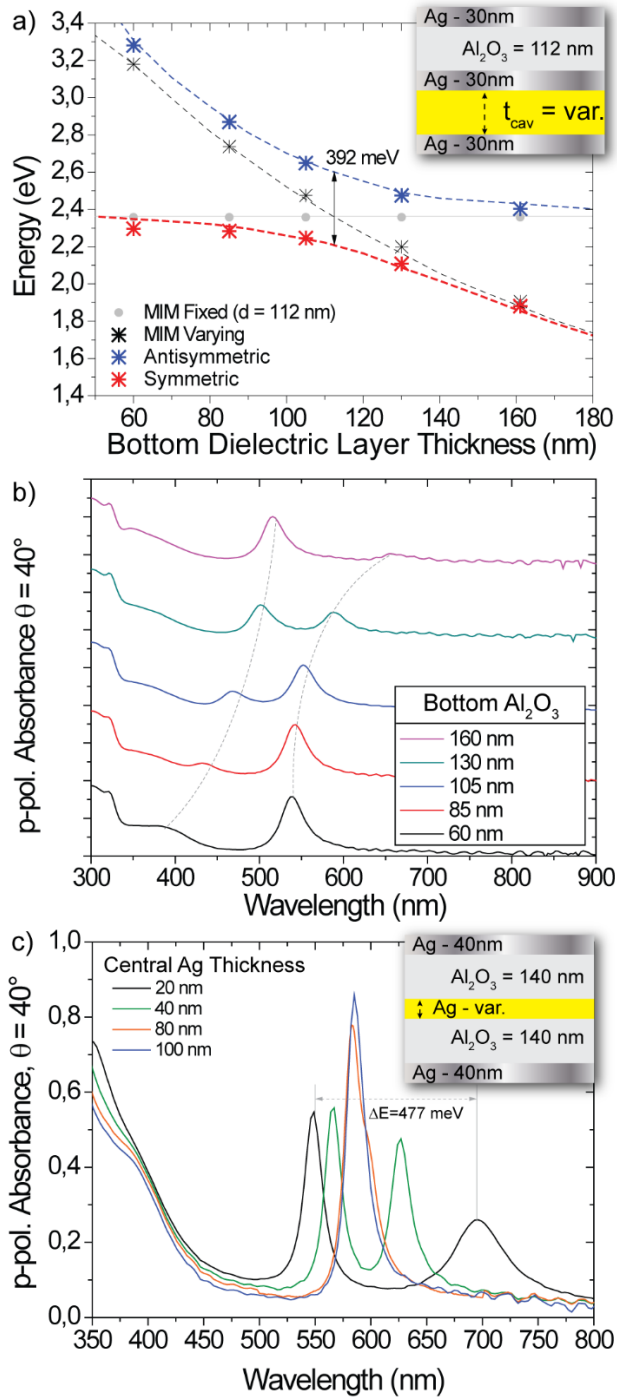

**Figure 2.** (a) Mode anticrossing in a MIMIM system with 30 nm Ag layers and 112 nm $Al_2O_3$ as top dielectric layer, while the thickness of the bottom dielectric layer is varied. Experimentally measured data is shown by stars, and the simulated dispersion via SMM by dashed lines. The grey

markers (experimental) and lines (SMM simulations) show the case of non-interacting cavities. (b) Absorbance spectra measured in p-polarization and obtained as (1-transmittance-reflectance) for the five MIMIM structures with different dielectric bottom layer thickness displayed in (a). (c) Ellipsometrically measured p-polarized absorbance curves for different thickness of the central layer. For a MIMIM with 20 nm central Ag layer we obtained a mode splitting of 447 meV that corresponds to about six times the linewidth of the antisymmetric mode (see also Figure S2 for MIMIM systems with other metal layer thicknesses).

Figure 2a shows the absorbance maxima of the high (blue stars) and low (red stars) energy resonances for 5 samples, where the bottom dielectric layer thickness was varied from 60 to 160 nm, while that of the top layer was constant at 112 nm. We clearly observe the anticrossing behavior that is expected for two coupled modes, corroborated by Scattering Matrix Method (SMM) simulations (dashed lines).[27,42,43] Careful inspection of the experimental and simulated frequencies in the anticrossing region reveals that the shift of the high energy (HE-ENZ) mode from the unperturbed frequency (shown by the solid grey line) is larger than that of the low energy (LE-ENZ) mode. Such behavior deviates from the classical coupled oscillator model where a symmetric mode splitting occurs. The corresponding spectra are shown in Figure 2b, where we notice that the resonance associated to the top layer manifests a more pronounced absorbance peak outside the strong anticrossing region. The coupling strength of the ENZ modes is determined by the thickness of the central metal layer, as evident from Figure 2c. Here, symmetric MIMIM cavities were fabricated with fixed dielectric and outer metal layer thicknesses, while the thickness of the central Ag layer was varied from 20 to 100 nm. Clearly the mode splitting decreases with increasing central layer thickness. The anticrossing behavior of MIMIM structures with different mode splitting is shown in Figure S2 in the SI. Another important factor for photonic cavities is the quality of the resonances in terms of linewidth and quality (Q) factor, where the latter is evaluated as the ratio of the full-width-at-half maximum over central resonance frequency. From the black spectrum in Figure 2c we obtain a Q-factor of 35 at 540 nm (2.3 eV) with a line width of 80 meV. However, such a performance can be noticeably improved by acting on the thickness of the external metal layers. The SMM calculations in Figure 3 demonstrate that the quality of the resonances is mainly determined by the thickness of the outer metal layers, while the mode splitting, determined by the central layer thickness, remains roughly constant. ENZ resonances with linewidth as low as 25 meV are achievable, together with very high Q factors of around 100 (Figure 4b) that correspond to plasmon relaxation times of the order of 200 fs.

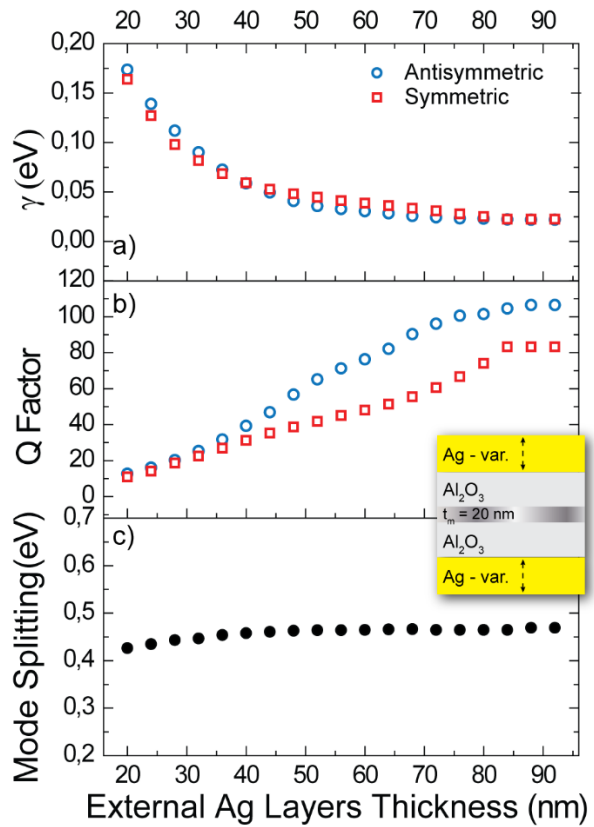

**Figure 3.** SMM simulations of the (a) linewidth, (b) quality factor, and (c) mode splitting of a MIMIM system with 20 nm thickness of the central metal layer, as a function of the thickness of the external metal layers.

The MIMIM structure is therefore a highly versatile photonic cavity, where the frequency of the resonance modes, their quality factor, and the mode splitting can be tailored to a very large degree throughout the visible and NIR spectral range. In particular, the spectral range of the resonances can be extended by an adequate choice of the dielectric material, as demonstrated in Figure S3 where the resonances for MIMIM cavities with $TiO_2$ and $SiO_2$ are shown. Interestingly, for $TiO_2$ also the asymmetric resonances manifest hybridization and mode splitting. Therefore, MIMIM double cavities are appealing when the spectral overlap with dyes, quantum emitters and other photonic systems is sought.

Motivated by the discrepancies of the optical properties of the MIMIM double cavity with the classical analytical model, we outline its analogy to a double quantum well in quantum mechanics that allows to resolve this problem. Furthermore, this treatment leads to analytical expressions for the cavity resonances that describe the MIMIM with high accuracy, and provides a more intuitive physical insight to such a complex photonic system. The geometry for the SMM calculation for MIMIM with thick external metal layers is shown in Figure 4a. The corresponding quantum well with infinitely thick external barriers is depicted in Figure 4b, together with the symmetric and antisymmetric eigenmodes that such a system sustains. In ref [36] we demonstrated that the square of the imaginary part of the dielectric permittivity of the metal, $\kappa$, can be seen as the optical equivalent of the potential that defines the barrier height. The semi-classical treatment for the MIM cavity that we developed on this basis can be straightforwardly extended to the MIMIM structure, thus allowing to find the dispersion relations that define the resonant modes:

$$tanh\left(k_0\kappa_m \frac{t_{cm}}{2}\right) = \frac{\kappa_m}{n_d} tan\left[-k_0 n_d \left(\frac{1}{k_0\kappa_m} + t_d\right)\right]$$

(2)

$$-coth\left(k_0\kappa_m \frac{t_{cm}}{2}\right) = \frac{\kappa_m}{n_d} tan\left[-k_0 n_d \left(\frac{1}{k_0\kappa_m} + t_d\right)\right] \qquad (3)$$

Details on the derivation of these relations are reported in the SI. Figure 4c shows the resonance frequencies obtained by solving eqs. 2 and 3 for different dielectric layer thicknesses, in the case of external metal layers with 100 nm thickness. The comparison with the numerical electromagnetic (SMM) simulations shows excellent agreement. Since we consider illumination of the structures from the top, *i.e.* through the metal layers, MIMIM structures with thinner Ag layers are of large practical interest. In this case, the tunneling of the photon through the barrier has to be taken into account by an additional phase factor of $-exp\left(-2k_0\kappa_m t_m\right)$. Then, the dispersion relation for a "leaky" MIMIM structure is:

$$tanh\left(k_0\kappa_m \frac{t_{cm}}{2}\right) = \frac{\kappa_m}{n_d} tan\left(-k_0 n_d \left(\frac{1}{k_0\kappa_m} + t_d\right)\right) - exp\left(-2k_0\kappa_m t_m\right) \qquad (4)$$

$$-coth\left(k_0\kappa_m \frac{t_{cm}}{2}\right) = \frac{\kappa_m}{n_d} tan\left(-k_0 n_d \left(\frac{1}{k_0\kappa_m} + t_d\right)\right) - exp\left(-2k_0\kappa_m t_m\right) \qquad (5)$$

The resulting resonance frequencies for the symmetric and asymmetric modes are plotted in Figure 4d as a function of the thickness of the external Ag layers. Here the thickness of the dielectric layers was fixed at 140 nm, and the central metal layer was 20 nm thick. We clearly observe a decrease in resonance frequency when the thickness of the external layers is smaller than 40 nm, which identifies the leaky regime, where the additional phase correction in eqs. 4 and 5 is necessary. Again, very good agreement with the numerical (SMM) modeling, shown by dashed lines, is obtained. The electric and magnetic field profiles of the resonant modes in such a leaky MIMIM structure, calculated by finite element method simulations (COMSOL Multiphysics), are reported in Figure S2 in the SI. From the experimental data in Figure 2c we already saw that the coupling strength of the modes is determined by the thickness of the central metal layer. The double quantum well model of the MIMIM illustrates this coupling as resonant tunneling of the photons between the two wells, and Figure 4e shows that also the asymmetric mode splitting is well described by the quantum mechanical treatment (depicted as small symbols). The experimental and theoretical data in Figure 4e demonstrate that the coupling in the MIMIM cavity can be tuned through the entire range, from uncoupled, to weakly coupled and strongly coupled by acting on the thickness of the central layer. The strong coupling regime is reached if $\frac{4g^2}{\gamma_{MIM_1} \cdot \gamma_{MIM_2}} > 1$ , where $g$ is the coupling constant ($g = \frac{\hbar\Omega}{2}$), with $\hbar\Omega$ being the mode splitting in meV, and $\gamma_{MIM_1}$ and $\gamma_{MIM_2}$ are the linewidths of the uncoupled modes. We note that for an infinitely thin central metal layer the resonances approach the symmetric and asymmetric modes in a MIM system, and that for a very thick central metal layer the bottom cavity is shielded by the top one, which leads to the optical response of a MIM superabsorber.[5,44] Figure 4 demonstrates that the semiclassical approach, treating the MIMIM as a double quantum well for photons, allows for accurate analytical calculation of the resonance energies, and provides physical insight in the mode coupling and mode confinement as related to photon tunneling through the metal barriers.

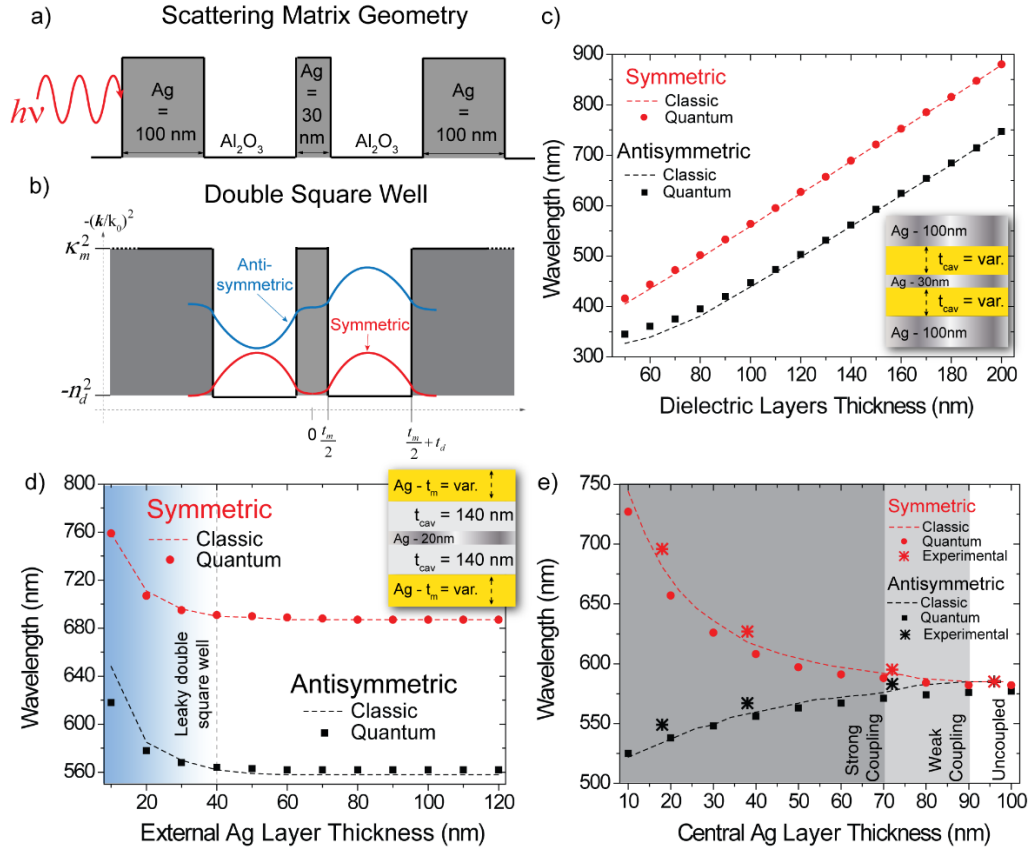

**Figure 4.** (a) Structure of a MIMIM with thick external metal layers as used for SMM calculations. (b) The double quantum well that is the quantum mechanical analogue of the MIMIM structure. The y-scale for the optical potential is - , and the origin of the x-axis is chosen in the middle of the central barrier. The barrier height of the potential for the photons induced by the metal layers is given by the square of the imaginary part of the refractive index $\kappa_m^2$ , while the potential in the dielectric is at $-n_d^2$. (c,d) Resonance frequencies calculated with the quantum approach (markers) and with numerical SMM simulations (dashed lines) for different thickness of the dielectric layers (c) and of the external Ag layers. (e) Experimental (large markers), quantum calculated (small markers), and SMM simulated resonance wavelengths of MIMIM structures with different central

metal layer thickness. The mode splitting can be tuned from uncoupled resonators, through weak coupling, into the strong coupling regime. The largest experimental mode splitting is 477 meV for a central metal layer thickness of 18 nm, which is among the highest values reported in literature.[12,13,53,45–52]

Conclusions

We demonstrated that the mode coupling in MIMIM double cavities allows to tailor their optical resonances in the visible spectral range. Spectroscopic ellipsometry identified the cavity resonances as ENZ modes, and a quantum mechanical view elucidated that their hybridization can be seen as resonant tunneling of photons through the central metal layer. Within the quantum mechanical treatment, we were able to describe the dependence of the spectral position of the ENZ modes on all geometrical and optical parameters of the constituent materials. MIMIM cavities can be a versatile platform for basic studies and practical applications where multiple tunable high-quality photonic resonances are needed.

Materials and Methods

*Fabrication and characterization of the MIMIM structures.*

A multistep process was followed to fabricate the MIMIM structures, consisting of deposition of (i) the metal (Ag), and (ii) the dielectric ($Al_2O_3$) stacked layers. (i) Ag layer were deposited via electron-beam induced thermal evaporation (Kurt J. Lesker PVD 75) on a glass substrate, followed by (ii) the deposition of 10 nm $Al_2O_3$ inside the same system, thus preventing Ag from oxidation. The remaining $Al_2O_3$ was deposited via Atomic Layer Deposition (ALD) (FlexAl from Oxford Instruments) using a thermal deposition process with a stage temperature of 110 °C, obtaining an alumina deposition rate of 0.09 nm/cycle. Tri-methylaluminate (TMA) and $H_2O$ were used as precursors. A heating step of 300 s was performed before starting the ALD cycles. Each ALD cycle consisted of a H$_2$O/purge/TMA/purge sequence with a pulse durations of 0.075/6/0.033/2 seconds, respectively.

Spectroscopic Ellipsometry was used to characterize the optical response and thickness of all the samples, as well as a versatile experimental tool for the detection of polarized scattering parameters (Reflectance, Transmittance and Absorbance). A Vertical Vase ellipsometer by Woollam was used, in the range from 300-900 nm. Spectroscopic analysis has been carried out at three different angles (50°, 60° and 70°) with a step of 3 nm. For both the experiments and spectroscopic characterization a resolution of 3 nm was selected, and all spectra have been normalized to the intensity of the Xe lamp.

*Modeling and Simulations.*

Finite Element Method based full-field simulations have been carried out by means of COMSOL Multiphysics. As boundary conditions, appropriately swept-meshed Perfectly Matched Layers have been imposed, while the nanometric layers constituting the MIMIM have been meshed with a free triangular texture in which the maximum element size is 2 nm. Illumination is provided by a plane wave impinging at 40° via a classic internal port excitation, adjacent to the upper Perfectly Matched Layer domain. Dielectric permittivities of all the selected materials were taken from experimentally measured data. Scattering Matrix Method simulations have been conducted by means of a customized MATLAB code.

ACKNOWLEDGMENTS

The research leading to these results has received funding from the European Union under the Marie Skłodowska-Curie Grant Agreement COMPASS No. 691185.